\documentclass[11pt,a4paper]{aa}
\usepackage[utf8]{inputenc}
\usepackage[T1]{fontenc}
\usepackage{lmodern}
\usepackage{microtype}
\usepackage{setspace}
\usepackage{graphicx}
\usepackage{booktabs}
\usepackage{enumitem}
\usepackage{hyperref}
\hypersetup{colorlinks=true,linkcolor=blue,citecolor=blue,urlcolor=blue}
\usepackage{titling}
\usepackage{fancyhdr}
\usepackage{amsmath,amssymb}
\usepackage{afterpage}
\usepackage{multicol}
\usepackage{xcolor}
\usepackage{bibentry}

\begin{document}

\begin{titlepage}
  \centering
  \vspace*{1cm}
  {\Huge\bfseries Expanding Horizons \\[6pt] \Large Transforming Astronomy in the 2040s \par}
  \vspace{1.5cm}

  {\LARGE \textbf{Nova Explosions in 2040}\par}
  \vspace{1cm}

  \begin{tabular}{p{4.5cm}p{10cm}}
    \textbf{Scientific Categories:} & Time-domain, spectroscopy, novae, cataclysmic variables\\
    \\
    \textbf{Submitting Author:} & Alessandro Ederoclite: \\
    & Centro de Estudios de Física del Cosmos de Aragón, Plaza San Juan 1, Planta 2, 44001, Teruel, Spain \\
    & \url{aederocl@cefca.es}\\
    \\
    \textbf{Contributing authors:} & 
    Domitilla De Martino$^1$,
    Paul Groot$^{2,3,4}$,
    Elena Mason$^5$,
    Gloria Sala$^{6,7}$,
    Martín Guerrero$^8$, 
    Thomas Kupfer$^{9,10}$, 
    Anna Francesca Pala$^{11}$,
    Simone Scaringi$^{12}$, 
    Noel Castro Segura$^{13}$
\\

\vspace{0.3cm}\\
\multicolumn{2}{l}{\scriptsize $^{1}$ INAF-OACNa, Salita Moiariello 16 80141 Naples,Italy}\\
\multicolumn{2}{l}{\scriptsize $^{2}$ South African Astronomical Observatory, PO Box 9, 7935, Observatory, South Africa}\\
\multicolumn{2}{l}{\scriptsize $^{3}$ Department of Astronomy, University of Cape Town, Private Bag X3, Rondebosch, 7701, South Africa}\\
\multicolumn{2}{l}{\scriptsize $^{4}$ Department of Astrophysics/IMAPP, Radboud University, PO Box 9010, 6500 GL, Nijmegen, The Netherlands}\\
\multicolumn{2}{l}{\scriptsize $^{5}$ INAF-OATS, Via G.B. Tiepolo 11, 34143, Trieste, Italy}\\
\multicolumn{2}{l}{\scriptsize $^{6}$ Departament de Física, EEBE, Universitat Politècnica de Catalunya, c/Eduard Maristany 16, 08019, Barcelona, Spain}\\ 
\multicolumn{2}{l}{\scriptsize $^{7}$ Institut d'Estudis Espacials de Catalunya (IEEC), c/ Esteve Terradas 1, 08060, Castelldefels (Barcelona), Spain}\\
\multicolumn{2}{l}{\scriptsize $^{8}$ Instituto de Astrofísica de Andalucía, IAA-CSIC, Glorieta de la Astronomía S/N, Granada, 18008, Spain}\\
\multicolumn{2}{l}{\scriptsize $^{9}$Department of Physics and Astronomy, Texas Tech University, 2500 Broadway, Lubbock, TX 79409, USA}\\
\multicolumn{2}{l}{\scriptsize $^{10}$Hamburg Observatory, University of Hamburg, Gojenbergsweg 112, 21029 Hamburg, Germany}\\
\multicolumn{2}{l}{\scriptsize $^{11}$ European Southern Observatory, Karl Schwarzschild Straße 2, D-85748, Garching, Germany}\\
\multicolumn{2}{l}{\scriptsize $^{12}$ Department of Physics, Centre for Extragalactic Astronomy, Durham University, South Road, Durham DH1 3LE, UK}\\
\multicolumn{2}{l}{\scriptsize $^{13}$ Department of Physics, University of Warwick, Gibbet Hill Road, Coventry, CV4 7AL, UK}\\

\vspace{0.5cm}
\textbf{Abstract:}
\vspace{0.5em}

  \begin{minipage}{0.8\textwidth}
    \small
Novae are thermonuclear explosions on the surface of accreting white dwarfs 
and are key laboratories for studying explosive nucleosynthesis, particle 
acceleration, shock physics, and binary evolution. Despite major progress 
driven by wide-field time-domain surveys and multi-wavelength facilities, 
our understanding of nova explosions remains limited by incomplete temporal 
coverage, heterogeneous spectroscopic follow-up, and poorly constrained 
ejecta properties. In this white paper we outline the open scientific 
questions that will define nova research in the 2040s, focusing on the mass, 
composition, geometry, and dynamics of the ejecta, the role of the underlying 
binary system, and the connection between nuclear burning, shocks, and emission 
across the electromagnetic spectrum. 
We argue that decisive progress requires rapid-response, high-cadence, 
multi-wavelength observations, anchored by systematic high-resolution 
optical and near-infrared spectroscopy from eruption to quiescence. 
Finally, we identify key technological requirements 
needed to enable transformative advances in the physics of nova explosions 
over the coming decades.
  \end{minipage}

  \end{tabular}
\setcounter{page}{0}

\end{titlepage}


\section{Introduction and Background}
\label{sec:intro}

Novae are thermonuclear explosions occurring on the surface
of a white dwarf which is accreting mass from a late-type companion. 
These events are invaluable laboratories for
the physics of thermonuclear reactions on compact objects, 
particle acceleration to TeV energies \citep{2014Sci...345..554A}, and they play a major role in the synthesis of some isotopic 
species (e.g. $^{15}$N $^{17}$O, \citealt{jordi}) and elements such as Lithium  (e.g.
\citealt{molaro}).
Whether the WD retains the accreting mass increasing 
up to Chandrasekhar limit and explode as a Type\,Ia supernova, which 
have a major role as cosmological standard candles, has also to be still understood.
For a review, see \cite{Chomiuk2021}.

\begin{figure*}
    \centering
    \begin{tabular}{c c}
       \includegraphics[width=0.45\linewidth]{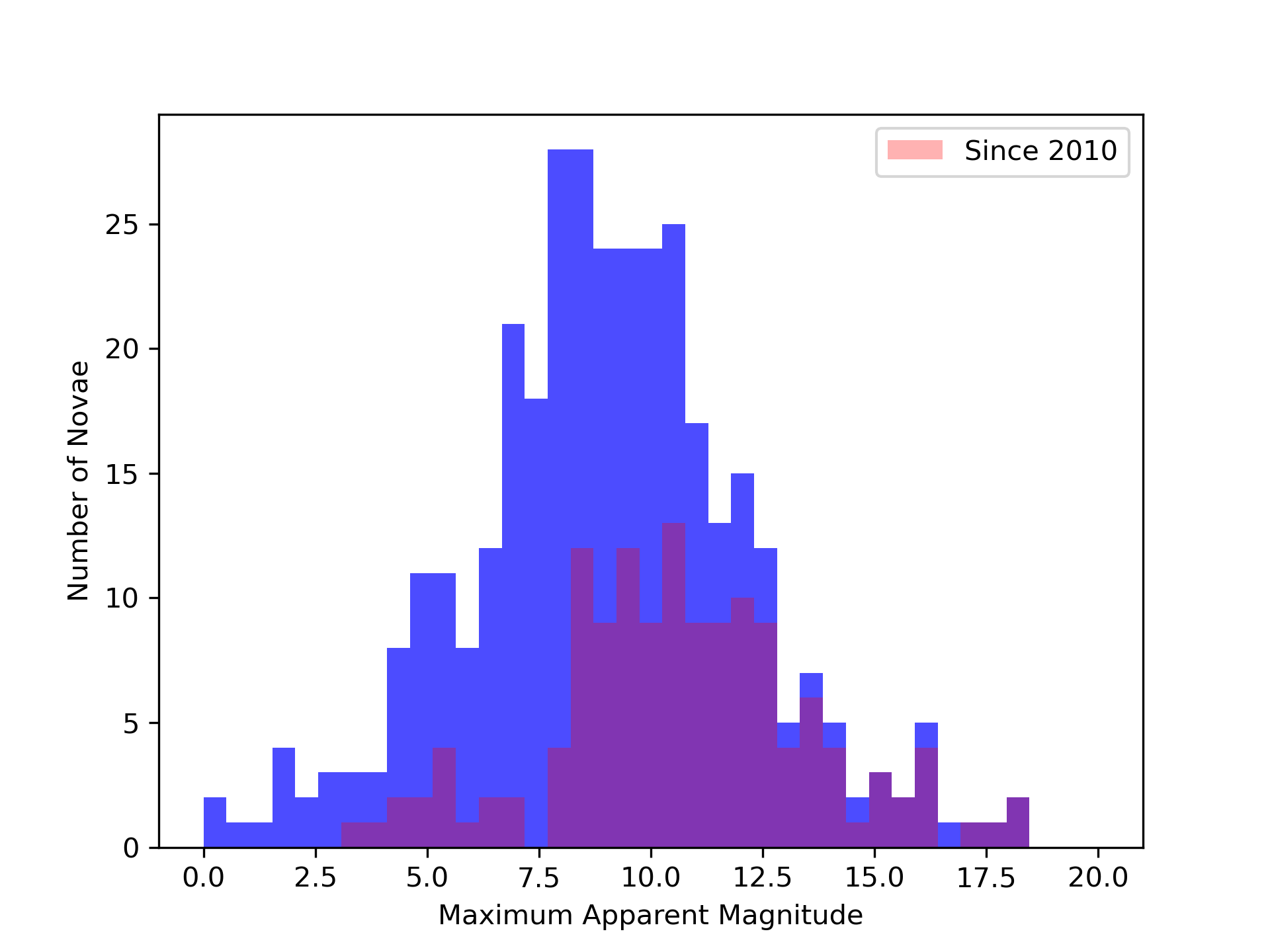}     &  
       \includegraphics[width=0.45\linewidth]{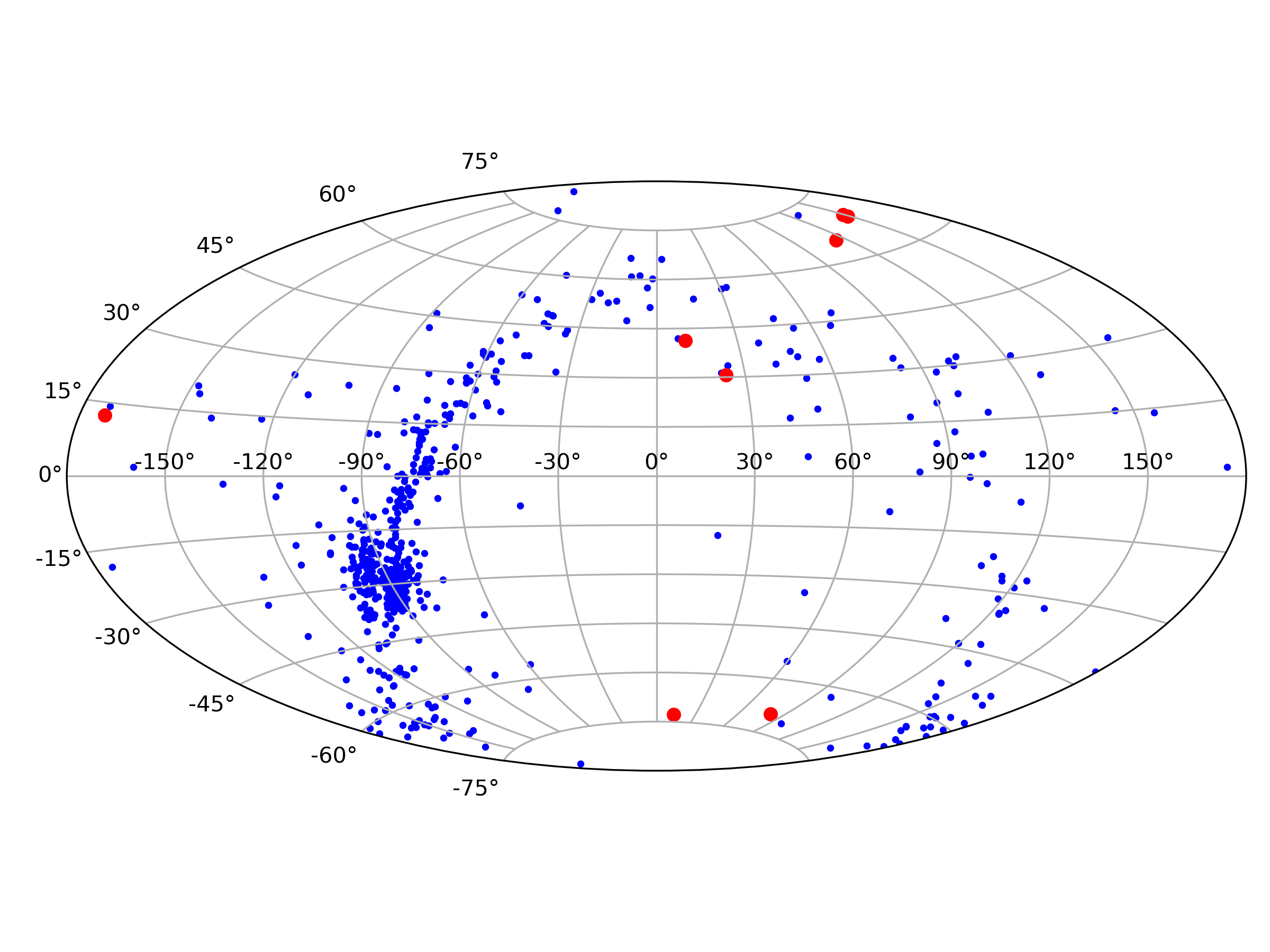}     \\ 
       \includegraphics[width=0.45\linewidth]{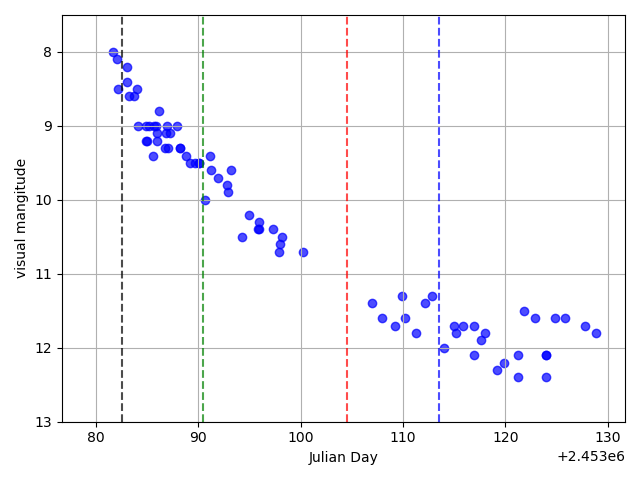}     &
       \includegraphics[width=0.45\linewidth]{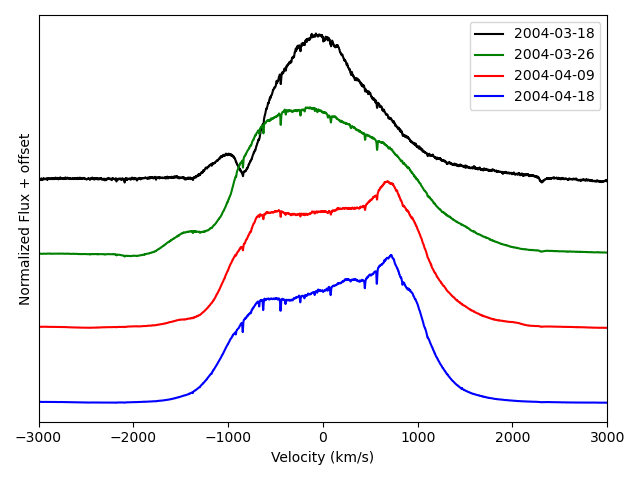}\\
    \end{tabular} 
    \caption{
    \textit{Upper Left:} Histogram of the magnitude at maximum
    of novae. In red, the histogram of the magnitude at
    maximum of novae discovered after 2010. For the purpose
    of this plot, the difference between photometric bands
    is neglected.
    \textit{Upper Right:} Location of Galactic novae (in blue) and
    galaxies which have spectroscopically confirmed novae (in red). 
    \textit{Lower Left:} AAVSO light curve of V5114\,Sgr. The 
    vertical lines mark the spectroscopic observations 
    obtained within the first 40 days.
    \textit{Lower Right:} High resolution (R$\sim$48,000) 
    spectroscopy of the nova centred on the H$\alpha$
    region. 
    }
    \label{fig:nova_statistics}
\end{figure*}

From an observational point of view, novae are seen
as the sudden increase in brightness of a point source
(histogram in the upper left panel of 
Fig.\ref{fig:nova_statistics}). 
Once considering the distance, novae at maximum have 
absolute magnitudes between $V\sim-6$ and $\sim-10$\,mag.
The sudden brightening is followed by a dimming whose rate
can vary. The time that a light curve needs to dim by $n$
magnitudes is called $t_n$. It is common to use $t_2$ or
$t_3$ as references for the evolutionary time scale of a nova.
In general terms, a nova is considered fast if $t_2<12$\,days
and slow otherwise \citep{V5114Sgr}.
The time required to reach a quiescent state is largely 
unknown, mostly because follow-up is rarely carried out
for more than a year.

There are about 10\,novae per year detected in the Milky Way (see the 
blue points in the upper right
panel of Fig.\,\ref{fig:nova_statistics}).
The Galactic nova rate has been estimated to range
between 10 and 300 novae per year (see \citealt{2021ApJ...912...19D}
and references therein).
Yet, they are rarely detected before maximum and a typical
spectroscopic follow up consists of a spectrum at maximum,
a handful of spectra between maximum and $t_3$ and then 
a handful of spectra during the
following year.
This is shown in Fig.\,\ref{fig:nova_statistics}: the light 
curve of V5114\,Sgr\,(2004) is from AAVSO 
and the spectra show the variation in shape of the H$\alpha$ line.
The number of extragalactic novae (see the red points in the 
upper right panel Fig.\,\ref{fig:nova_statistics}) 
being detected
is increasing thanks to projects like ZTF and BlackGEM.

Over the past two decades, wide-field time-domain 
photometric surveys have
provided critical insights on the properties of novae.
In particular, the change from a discovery mostly relying on 
amateur astronomers to automatic detection of transients have
increased the number of detection of fainter novae (see the 
upper left panel of Fig.\ref{fig:nova_statistics}).
\cite{hounsell} have used the Solar Mass Ejection Imager (SMEI) on 
board the Coriolis satellite and derived light curves with a
102\,minutes cadence. These light curves have an unprecedented quality, 
in particular in the pre-maximum region.
Arguably, some
of the most complete light curves currently come from networks of telescopes
with nodes covering various continents (e.g. ATLAS and ASAS-SN).
Yet, these light curves are normally limited to magnitudes 
lower
than 18, hence hardly ever capturing the full extent of the 
nova phenomenon.

Spectroscopic studies have produced significant advancements 
in the physical characterization of the nova events thanks to 
prolonged  monitoring (up to few year after outburst) through 
high resolution spectroscopy 
but only for a handful of novae 
\citep{shore2012,deGennaroAquino+2014,mason+2018,shore+2018}.   
The two largest spectroscopic databases are the Stony Brook SMART Atlas and 
the collection from the amateurs of the ARAS group, often 
limited in resolution or the monitoring duration because 
of the available collecting power.  
Further progress in the nova understanding requires more 
systematic observations in high resolution spectroscopy and
improved cadence.
Novae have long been thought to come in two different 
populations, named after the main non-hydrogen lines
observed at maximum light: ``He/N'' and ``Fe\,II''.
Thanks to high resolution and high cadence observation before
and after maximum light, \cite{elias} demonstrated that
the two classes were a by-product of the bias due to the
sub-optimal temporal spectroscopic coverage of the nova
evolution.
Nova explosions power bright emission at all wavelengths, from radio to VHE with energies greater than 100 GeV \cite{2010Sci...329..817A, 2014Sci...345..554A, 2022Sci...376...77H}. The production of VHE emission by proton acceleration was, in fact, first proposed for RS Oph \cite{2007ApJ...663L.101T} and was confirmed when Fermi detected $\gamma$-ray emission in its last outburst \cite{2022ApJ...935...44C}. The thermal radio emission from the shocks of the expanding shell with a dense external shell and their relation to the VHE emission was modeled by \cite{2014MNRAS.442..713M},  showing that at early times when the forward shock is radiative, radio emission originates from a dense cooling layer immediately downstream of the shock. X-rays probe simultaneously the H-burning power source at the white dwarf surface and the expanding and shocking ejecta. IR emission in novae probes dust and warm gas heated by radiation and shocks, while radio emission is powered by ionized ejecta (free–free emission) and, in many cases, shock-driven synchrotron radiation. However, the link between ejected matter emitting at  different energies 
has still to be assessed due to the lack of appropriate monitoring timing.


The examples from \cite{hounsell,elias,farung} highlight that the study of
novae requires: (i) quick response, (ii) high cadence, (iii) multi-wavelength, 
(iv) high spectral resolution, (v) from the onset of the explosion until the
return to quiescence.


\subsection{The Opportunities for the Near Future}


The Large Survey of Space and Time performed at the Vera Rubin Observatory
is one of the largest projects of the decade. Although the cadence of the
observations is not optimal for nova studies, the depth of the observations 
is such that several binaries hosting a nova explosion may be within 
reach.

The ``Son of X-Shooter'' (SOXS) is to be installed on the 3.5\,m New Technology
Telescope at La Silla Observatory. 
Covering the whole optical and
near-infrared in a single observation 
at a modest spectral resolution and on a mid-sized (yet dedicated) 
telescope, it will be the first dedicated facility to address the 
long-sought time domain in spectroscopy mode.

The Extremely Large Telescope will 
provide angular 
resolution of a few milliarcseconds (depending on the wavelength). 
Combined with the high spectral resolution
in the near infrared provided by HARMONI, this is promising
but limited to address the multi-variate 
spectroscopic behaviour of novae (e.g. 
\citealt{shafter+2011, shore_tpyx, wms2012}).

The Square Kilometer Array (SKA) is a radiotelescope made of antennas 
spread over Australia and South Africa covering from 50 MHz to 14 GHz.
The large amount of antennas gives it flexibility to react
to transient phenomena
tracking the evolution of ejected 
matter and jet formation (e.g. \citealt{sokoloski_2008}).

The Cherenkov Telescope Array (CTA) is a ground-based observatory using
Cherenkov radiation to observe at TeV energies from both hemispheres.
Close-by novae can be detected with current (MAGIC, HESS) instrumentation and 
the higher sensitivity and whole sky coverage of CTA means the possibility
to investigate a larger number of novae.

In the decade of the 2040s, X-IFU on board
NewATHENA will probe all phases at unprecedented spectral
resolution: from early hard X-rays, which pinpoint
the presence of the evolving shocked ejecta (see  
\citealt{2015JHEAp...7..117O}), to the luminous super-soft 
component directly probing the main power source of the
post-nova (i.e., the H-burning white dwarf atmosphere). With
the launch planned in 2030, the wide-field camera W2C on-board 
eXPT will be able to detect the early X-ray emissions
from bright novae, while the SFA will then track the X-ray
spectral evolution, and PFA test the X-ray polarimetry 
\citep{2025SCPMA..6819507Z}. 
However, the full picture of the interplay of
H-burning power source (which depends on the WD mass)
and the physics of the ejecta (shocks, energetics, ionization,
and cooling processes) can only be obtained together with
a regular optical spectroscopic follow-up, which traces the
abundances, ionization, and expanding velocities of the ejecta.

GaiaNIR, if approved, may become the
NIR counterpart of {\it Gaia}. Its catalogue would be deeper than its
optical counterpart in the plane of the Milky Way (where novae are more
numerous, see Fig.\,\ref{fig:nova_statistics}) and the 
astrometric solution 
would improve considering the almost 20\,years baseline.

\section{Open Science Questions in the 2040s}
\label{sec:openquestions}

\subsection{Mass and Composition of the Ejecta}

The least constrained parameter in nova observations is the amount of mass 
ejected in an explosion. 
There are three main sources of uncertainty: the distance to the
nova, the interstellar extinction and the clumpiness of the ejecta.

The measure of the distance of Galactic novae is a notoriously difficoult task.
Only one nova could be measured with light echo (T Pyx, \citealt{tpyx_light_echo}), a handful of novae could be determined
with expansion parallax and less than 10\% of known novae have distances which can be
inferred from {\it Gaia} \citep{nova_gaia_distances}. 

The interstellar extinction towards a nova is crucial since most novae occur 
in the plane of the Milky Way and can be observed as far as 8\,kpc.
In this case, the use of several and 
complementary probes of interstellar extinction are necessary (e.g., see
\citealt{V5114Sgr, shore_tpyx}). In the case of ground based observations,
this requires access to the whole range from 300-2400\,nm at spectral resolution
$\geq$40,000.

Several direct observations show that nova ejecta 
are clumpy (e.g. \citealt{schaefer_tpyx,tiina}). 
So far,  
attempts to provide a direct spectroscopic probe of
clumpiness has been made by \cite{wms1994}
using  the $[$OI$]$ $\lambda\lambda6300-6364\AA$ doublet
and \cite{mason+2018} using the optically thin H$\beta$ emission.

\subsection{The Dynamics of the Ejecta}

Various authors have described the expansion of
nova ejecta as bipolar (e.g. \citealt{tpyx_Chesneau,shore2013}).

Current AO-systems on ELT-sized telescopes can achieve spatial 
resolutions down to 10 milliarcseconds in the infrared. 
Near-infrared interferometry (e.g. GRAVITY at VLTI) can 
achieve an angular resolution 5 milliarcseconds. 
Integral field spectroscopy with high spectral resolution
and this kind of spatial resolution would provide the 
evolution of the physical structure of the nova ejecta.
Similar work has been done by \cite{takeda,guerrero+2025}.

Optical and near-infrared interferometry can provide 
insights but they are limited by the spatial resolution.
Linear spectropolarimetry is, in principle, only 
limited by the signal-to-noise and it can probe asymmetries
in the very early stages of the evolution of a nova, even 
when the resolution with interferometry or AO is prohibitive,
thus becoming a powerful complementary tool. 
Unfortunately, the use of this
technique is very limited in this field (e.g. \citealt{ikeda}).


\subsection{The Role of the Underlying Binary}

The metallicity of the material accreted on the white dwarf must have
a direct impact on the properties of the explosion. Yet, determining the
atmospheric properties of the secondary star in a CV which hosted a 
nova explosion is normally difficult as the accretion disc outshines 
the rest of the system even at infrared wavelengths.

The use of extragalactic novae can be a solution to this aspect. 
\cite{kasliwal}, using extragalactic novae, has identified a 
class of novae which look under-luminous with respect to the 
average brightness of novae with similar time scales in the Galaxy.
The reported spectroscopic analysis shows no difference between
Galactic ``ordinary'' fast novae and ``faint and fast'' nova discovered
in this work. Yet, the spectra of these novae are merely 
a confirmation of the nature of the object. Full high-cadence
spectroscopic follow-up would be required to be able to tell if
there are physical differences between bright and faint fast novae
in stellar populations with different metallicities.




The binary system conditions after a nova outburst are not known nor well investigated (observationally and theoretically). Nevertheless the time a WD takes to relax and accrete again in degenerate conditions drives the onset of the next nova outburst, whether it occurs in the next 1 or 10$^4$ years. 
\cite{mason+2021} suggest that nova Cen 2013 WD was not relaxed yet and bloated 5 years after the outburst. They did not find evidence of an accretion disk either. It should be verified whether this depends on the amount of leftover envelope and its cooling time since it might be directly connected to the ultimate fate of the WD mass (growing toward the Chandrashekar limit or not). 
Increasing the number of observed aftermaths and their evolution in time is critical to address this question.


\section{Technology and Data Handling Requirements}
\label{sec:tech}



In order to perform studies which will allow us to perform 
transformative studies in the field of novae, there are five
main technical requirements:
\begin{itemize}
    \item high spectral resolution (larger than 50,000) covering the whole 
    optical and near infrared range (from 300 to 2400\,nm) in a single 
    observation
    \item linear spectropolarimetry with moderate spectral 
    resolution (larger than 50,000) covering the region 
    between 350 and 700nm in a single observation
    \item integral field spectroscopy with the same spectral resolution as above and
    spatial resolution of 1\,milliarcsecond.
    \item abiliity to take advantage of the aforementioned spectral resolutions
    over a wide range of magnitudes; this could be achieved with telescopes of
    different sizes or changing the number of telescopes feeding the same 
    instrument
    \item possibility to schedule observations every day from the onset of the 
    explosion until the source goes back to quiescence;
   ideally
    in both hemisphere
\end{itemize}




\bibliographystyle{aa}

\bibliography{refs}  

\end{document}